\documentclass[structabstract]{aa}

\usepackage{graphicx}
\usepackage{amsmath}
\usepackage{epsfig}
\usepackage{txfonts}

\begin{document}

   \title{Spectral modeling of circular massive binary systems}

   \subtitle{Towards an understanding of the Struve--Sahade effect?}

   \author{M. Palate
          \inst{1}
          \and
          G. Rauw
          \inst{1}
          }

   \institute{Institut d'Astrophysique et de G\'eophysique, 
   Universit\'e de Li\`ege, Bât. B5c, All\'ee du 6 Ao\^ut 17, 4000 Li\`ege, Belgium\\
             \email{palate@astro.ulg.ac.be}
             }

   \date{Received <date>; accepted <date> }

	\abstract
   {Some secondary effects are known to introduce variations in spectra of massive binaries. These phenomena (such as the Struve--Sahade effect, difficulties to determine properly the spectral type,...) have been reported and documented in the literature.}
   {We simulate the spectra of circular massive binaries at different phases of the orbital cycle and accounting for the gravitational influence of the companion star on the shape and physical properties of the stellar surface.}
   {We use the Roche potential to compute the stellar surface, von Zeipel theorem and reflection effects to compute the surface temperature. We then interpolate in a grid of NLTE plan-parallel atmosphere model spectra to obtain the local spectrum at each surface point. We finally sum all the contributions (accounting for the Doppler shift, limb-darkening, ...) to obtain the total spectrum. The computation is done for different orbital phases and for different sets of physical and orbital parameters.}
   {Our first models reproduce the Struve--Sahade effect for several lines. Another effect, surface  temperature distribution is visible but the distribution predicted by our current model is not yet consistent with observations. }
   {In some cases, the Struve--Sahade effect as well as more complex line intensity variations could be linked to blends of intrinsically asymmetric line profiles that are not appropriatly treated by the deblending routine. Systematic variations of lines for (nearly) contact systems are also predicted by the model.}


   \maketitle

\section{Introduction}

	Massive stars are rare but very important objects. Their strong stellar winds interact with the ambient interstellar medium and can trigger the formation of new 	stars. They also play a key role in the production of various chemical elements. However, many open questions remain on these stars, relating to their formation, their evolution, their physical properties... One of the best ways to determine physical properties (such as masses, temperatures, radii ...) of massive stars is to study eclipsing binary systems. However, to get more information, we need to improve our models and take into account effects linked to the peculiarity of massive binary systems. Several authors (Sana et al. \cite{Sana}, Linder et al. \cite{Linder}, Linder \cite{LinderPhDTh}, and references therein) presented evidence of the impact of such peculiar effects on the spectral classification and on the spectra themselves. Such effects introduce complications to assign proper spectral types and luminosity classes. For example, main sequence stars are seen as giant stars (e.g. CPD-41\degr7742, Sana et al., \cite{Sana}). As the spectral type of a star is one of the most important pieces of information that we can obtain from observation, it is crucial to have reliable criteria of classification for both isolated stars and binary systems. One of the most important phenomena that impact the spectral classification is the Struve--Sahade effect that consists in a variation of the relative intensity of some spectral lines of the secondary with respect to the primary star lines with orbital phase (Struve, \cite{Struve}).  Another important effect is the surface temperature distribution that could affect the radial velocity determination. Indeed, on a heavily deformed star, the temperature at the stellar surface is not uniform and the lines are not all formed at the same place. This leads to different radial velocity determinations as a function of the line that we use.
	
	This work presents a novel way of simulating the spectra of binary systems at different phases of the orbital cycle. In a first approach, we model the case of binary systems with circular orbit and containing main sequence stars. First, we use the Roche potential to compute the shape of the stars. Then, we apply the von Zeipel (\cite{VonZeipel}) theorem and reflection effect to obtain the local temperature at the stellar surface. Finally, we use these geometrical information to compute synthetic spectra from a grid of model atmosphere spectra.
	
	Sect. 2 describes the assumptions, the modeling of the geometry of the star and the modeling of the spectra. We also present the important issues of the computation and the way we have treated them. Sect. 3 gives our first models and presents detailed analyses of the spectral classification, the Struve--Sahade effect, the light curves and radial velocity curves. We provide a summary of our results and the future perspectives in Sect. 4.	

\section{Description of our method}
	
	The modeling of circular binary systems can be done in following a quite simple approach that we can split in two parts: on the one hand, the surface, gravity and	temperature calculation and on the other hand, the spectra calculation. We develop here these two steps.
	
	\subsection{Geometric modeling}
	 	First of all, we need to calculate the surface of the stars of the binary system. Throughout this paper, we make the assumption of a circular orbit with both stars co-rotating with the binary motion, and acting like point-like masses. Under these assumptions, the stellar surface is an equipotential of the classical Roche potential (see e.g., Kopal, \cite{Kopal}):
	
		\begin{multline}
			\phi=-\dfrac{Gm_{1}}{a\sqrt{x^{2}+y^{2}+z^{2}}}-\dfrac{Gm_{2}}{a\sqrt{(1-x)^{2}+y^{2}+z^{2}}}\\
			-\dfrac{G(m_{2}+m_{1})}{2a}\left(x^{2}+y^{2}\right)+\dfrac{Gm_{2}x}{a} \nonumber
		\end{multline}
			
		where $x$ is the axis running from the center of the primary towards the secondary, $z$ the axis from the center of the primary and perpendicular to the orbital plane, and $y$ is perpendicular to $x$ and $z$. The $x$, $y$ and $z$ coordinates are dimensionless. $m_1$ and $m_2$ are respectively the masses of the primary and the secondary, $a$ is the separation between the stars and $G$ is the constant of gravity
			
		This can be written in adimensional form using spherical coordinates centered on the 									primary's center of mass:
		
		\begin{equation*}
		 \begin{array}{r c l}
				\Omega 	& = & -\dfrac{a\phi}{Gm_{1}} \\ [3mm]
						 		& = & 																																																\dfrac{1}{r}+\dfrac{q}{\sqrt{r^{2}-2r\cos\varphi\sin\theta+1}}+\dfrac{q+1}{2}\cdot
						 		r^{2}\sin^{2}\theta\\
						 		&& \multicolumn{1}{r}{-qr\cos\varphi\sin\theta}\\
		 \end{array}
		\end{equation*}
				
		where $r=\sqrt{x^{2}+y^{2}+z^{2}}$, $q=\frac{m_{2}}{m_{1}}$, $x=r\cos\varphi\sin\theta$, $y=r\sin\varphi\sin\theta$, $z=r\cos\theta$.
		We represent the stellar surface by a discretized grid of $240\times60$ points (in $\theta$ and $\varphi$ respectively). For a given value of $\Omega$, $r(\Omega, \theta, \varphi)$ is obtained by iterative solution of this equation using a simple Newton-Raphson technique. The knowledge of $r(\Omega, \theta, \varphi)$ over the discretized stellar surface allows us to calculate the gravity. Indeed, the local acceleration of gravity is given by the gradient of the Roche potential: 
		
		\begin{equation*}
			\underline{\nabla}\Omega=\left(\begin{array}{c}
					\dfrac{\partial\Omega}{\partial x}\\[3mm]
					\dfrac{\partial\Omega}{\partial y}\\[3mm]
					\dfrac{\partial\Omega}{\partial z}\\[3mm]
					\end{array}
				\right)=
				\left(
				\begin{array}{c}
					-\dfrac{x}{r^3}+\dfrac{q\left(1-x\right)}{r'^3}+\left(q+1\right)x-q\\[3mm]
					-\dfrac{y}{r^3}-\dfrac{qy}{r'^3}+\left(q+1\right)y\\[3mm]
					-\dfrac{z}{r^3}-\dfrac{qz}{r'^3}
				\end{array}
				\right)
		\end{equation*}
		
		where $r=\sqrt{x^{2}+y^{2}+z^{2}}$ as above, and \\
		$r'=\sqrt{(1-x)^{2}+y^{2}+z^{2}}$
		
		\begin{figure}[!ht]
			\resizebox{\hsize}{6cm}{\includegraphics{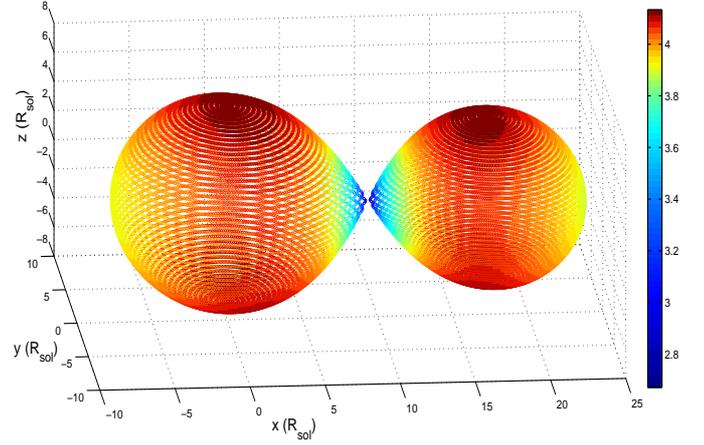}}
			\caption{Example of variations of log(g) for our Model 3.}
			\label{fig:log(g)Model3}	
		\end{figure}
			
		If we assume that we know the temperature at the pole, we can calculate the local temperature of each surface point following the von Zeipel (\cite{VonZeipel}) theorem:		
		
		\begin{equation*}		T_{\mathrm{local}}=T_{\mathrm{pole}}\left(\dfrac{\left\Vert\underline{\nabla}\Omega_{\mathrm{local}}\right\Vert}{\left\Vert\underline{\nabla}\Omega_{\mathrm{pole}}\right\Vert }\right)^{0.25g}			
		\end{equation*}
		where $g=1$ in the case of massive stars.
		
		We have explicitly made the assumption of a co-rotating system, so that the stars always present the same face to each other. This leads to a local increase of temperature due to the reflection effect between the two stars. We follow the approach of Wilson (\cite{Wilson}) to treat this effect. This approach consists of an iterative process in which we assume a radiative equilibrium on each point of the surface and calculate a coefficient of reflection. Once we reach the convergence, we update the temperature following: 
	
		\begin{equation*}
			T_{\mathrm{new}}=T_{\mathrm{old}}\cdot \mathcal{R}^{0.25}
		\end{equation*}
		with $\mathcal{R}$ the coefficient of reflection.
		
		\begin{figure}[ht!]
			\resizebox{\hsize}{6cm}{\includegraphics{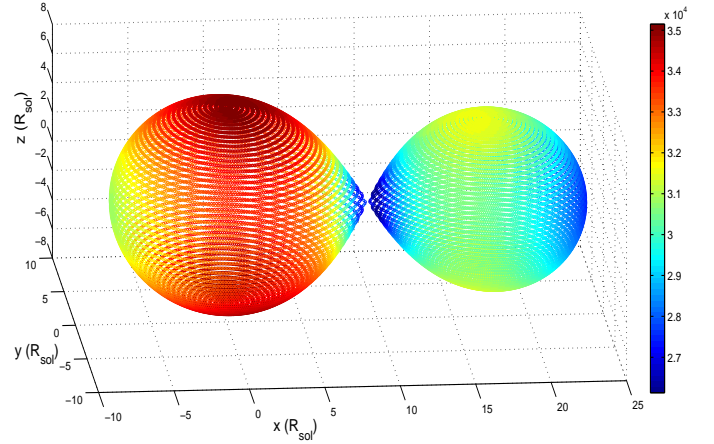}}
			\caption{Example of variations of temperature for our Model 3.}
			\label{fig:tempvarModel3}
		\end{figure}
		
	\subsection{Spectral modeling}
	
		The second part of the algorithm computes the spectrum of the binary by summing the incremental contributions of each surface point. We use a grid of OB star synthetic spectra computed with TLUSTY (Hubeny \& Lanz, \cite{Hubeny}) assuming solar metallicity to create a database of spectra (Lanz \& Hubeny, \cite{Lanza}, \cite{Lanzb}). Each spectrum is defined by two parameters: gravity and temperature. As we know these parameters for each point at the stellar surfaces, we can compute the local spectrum. The computation consists of a linear interpolation between the four nearest spectra in the grid. The next step consists in applying the appropriate Doppler shift to the spectrum accounting for the orbital and rotational velocity (the latter under the assumption of co-rotation) of the surface element.
	
		We multiply the spectrum by the area of the element projected on the line of sight towards the observer. The 	last corrective factor applied to the local contribution to the spectrum is the limb-darkening.	The  limb-darkening coefficient is based on tabulation by Al-Naimiy (\cite{Al-Naimiy}) for a linear limb-darkening law. Finally, we sum the contribution to the total spectrum. Note that we assume that there is no (or little) cross-talk between the different surface elements as far as the formation of the spectrum is concerned.

		 \begin{figure*}
				\sidecaption
  			\includegraphics[width=12cm]{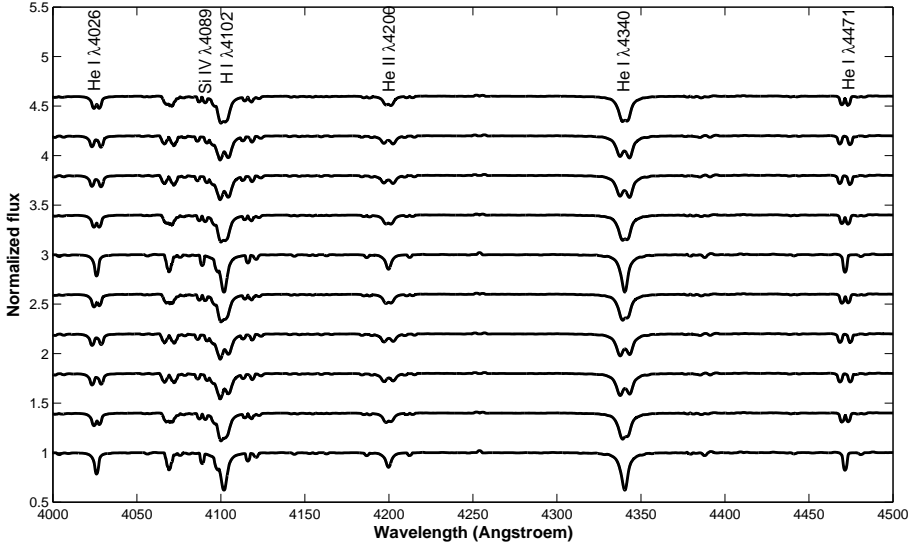}
  			\caption{Example of normalized simulated spectra computed for our Model 1. From bottom to top: phases $0$ to $0.9$ by step of $0.1$, the spectra are shifted vertically by $0.4$ continuum units for clarity.}
     	  \label{fig:spectra}
		\end{figure*}

		It is important to mention that we compute the spectra at different phases. Indeed, only one ``hemisphere" of each star is visible for an observer at any given time. However, as the stellar surface is not a sphere, the surface gravity and effective temperature of the surface elements visible at any given phase change with orbital phase. This leads to variations of the mean visible temperature of the stars and thus to variations in the line properties. Furthermore, when the primary shows its hottest part, the secondary shows its coolest part. Therefore, it reinforces the contrast between the spectra of the two stars. In the following, phase zero corresponds to the eclipse of the primary by the secondary. Over the first half of the orbital cycle (phase = $[0,0.5[)$, the primary star has a negative radial velocity. It has to be emphasized that our simulations allow us to compute the individual spectra of each component at each phase. This information is not available for real binary systems, where disentangling can provide average individual spectra, but usually does not allow us to derive individual spectra for each phase. In this first version, we do not yet take into account radiation pressure of the star that can possibly modify the shape of the atmosphere with respect to the Roche potential of very luminous stars neither the possibility of eccentric systems. These improvements will be done in a subsequent paper.

	\subsection{Computational issues }

		For the contact binaries, some problems occur at the contact point (L1 point). If we follow the von Zeipel (\cite{VonZeipel}) theorem, the temperature normally drops to zero because the gravity is null at the L1 point. However, on the one hand, such a result is not very physical. On the other hand, numerical instabilities occur in the Newton-Raphson scheme for stars filling their Roche lobe at positions near L1. These instabilities affect the determination of the local radius and surface gravity. In order to avoid this problem, we do not follow the equipotential strictly passing through the L1 point but an equipotential passing near this point. Another problem comes from the lack of some synthetic spectra in the model grid. This can occur, for instance, when a high temperature is coupled with a low gravity. The algorithm treats such problems by interpolating only between the two nearest spectra in the grid and notifies the user. However, this kind of situations occurs only for very few points located near the L1 point, which have a low temperature and have a very small contribution to the total stellar spectrum. These points generally do not influence significantly the global result. Another surprising problem comes from reflection. In the highly deformed binary, a coupling of some points near L1 leads to a high increase of temperature. At these positions, the convergence of the reflection computation is not guaranteed. In the approach of Wilson (\cite{Wilson}), the convergence is evaluated for each point and the iterative procedure is stopped once all the points have converged. We modify this global convergence criterion to ensure the end of the iteration process. The new criterion is to have at least $99$\% of the points converged. In some cases, despite this new criterion, some very few points present an abnormal temperature unrelated to that of of the surrounding points. These points are located near the L1 point and no spectra in the grid are available for these combinations of gravity and temperature. 
		
\section{First models }

	The number of well-known massive O-type binary systems with a circular orbit is rather small. Thus, we have selected five well studied massive binaries to test our algorithm and try to model the effects of binarity on the spectrum. The five systems are: HD\,159176, HD\,165052, HD\,100213, CPD-41\degr7742 and AO Cas. These systems have been studied by Linder et al. (\cite{Linder}, paper I) for the first three, Linder (\cite{LinderPhDTh}, paper III) for the fifth and Sana et al. (\cite{Sana}, paper II) for the fourth respectively. For the first three systems, paper I reports a Struve--Sahade effect, the last two do not seem to present this effect. These systems are main-sequence systems of massive OB-stars except AO Cas in which the secondary is a giant star. Table\,\ref{tab:*PARAM} lists the adopted parameters for each star. The parameters come from the observational analysis carried out in papers I, II and III. However, some systems (HD\,159176, HD\,165052) are not eclipsing and some parameters are not available. To overcome this problem, we use the calibration of Martins, Schaerer \& Hillier (\cite{Martins}) to assign a ``typical" value of the radius and mass. Hence, our models do not strictly correspond to the five observed systems. To avoid any confusion, HD\,159176, HD\,165052, HD\,100213, CPD-41\degr7742 and AO Cas models will be named respectively Model 1, 2, 3, 4 and 5.
	The first step of the computation is the surface, gravity and temperature evaluation (see Tables \ref{tab:TR} and \ref{tab:Gravity}). With these results, we can compare the simulated local gravity to the classical value for a single star: $g=\tfrac{Gm_{\mathrm{star}}}{R_{\mathrm{star}}^{2}}$.
			
	\begin{table*}[ht!]
		\caption{Parameters of the binary systems simulated in this paper. Inclinations in brackets stand for non eclipsing systems.}
		\label{tab:*PARAM}
		\centering
			\begin{tabular}{l c c c c c}
				\hline\hline 
				Parameters & Model 1 & Model 2 & Model 3 & Model 4 & Model 5\\
				\hline
				Period (day) & $3.36673$ & $2.95515$ & $1.3872$ & $2.44070$ & $3.52348$\\
				Mass ratio & $0.96$ & $0.87$ & $0.68$ & $0.55$ & $0.62$\\
				Semi-major axis ($R_{\sun}$) & $38.23$ & $31.25$ & $17.34$ & $23.14$ & $28.57$\\
				Inclination (\degr) & $(48)$ & $(23)$ & $77.8$ & $77$ & $65.7$\\
				Mass of primary ($M_{\sun}$) & $33.8$ & $25.15$ & $21.7$ & $17.97$ & $15.59$\\
				Mass of secondary ($M_{\sun}$) & $32.41$ & $21.79$ & $14.7$ & $9.96$ & $9.65$\\
				Primary polar temperature (K) & $38000$ & $35500$ & $35100$ & $34000$ & $33700$\\
				Secondary polar temperature (K) & $38000$ & $34400$ & $31500$ & $26260$ & $30500$\\
				$\Omega_{L1}$ & $3.68$ & $3.53$ & $3.20$ & $2.98$ & $3.10$\\
				\hline
			\end{tabular}
	\end{table*}

	\begin{table*}
		\caption{Temperature and radius of stars.}
		\label{tab:TR}
		\centering
			\begin{tabular}{l c c c c c c}
				\hline\hline
				Stars & $R_{\text{min}}$ & $R_{\text{max}}$ & $R_{\text{mean}}$ & $T(R_{\text{min}})$ &
				$T(R_{\text{max}})$ & $T_{\text{mean}}$ \\
				 & ($R_{\sun}$) & ($R_{\sun}$) & ($R_{\sun}$) & (K) & (K) & (K) \\
				\hline	
				Model 1$^{\text{1}}$ & $9.37$ & $9.79$ & $9.49$ & $38008$ & $37328$ & $37617$\\
				Model 1$^{\text{2}}$ & $8.94$ & $9.31$ & $9.05$ & $38008$ & $37516$ & $37664$\\
				Model 2$^{\text{1}}$ & $9.10$ & $9.82$ & $9.29$ & $35520$ & $34178$ & $34896$\\
				Model 2$^{\text{2}}$ & $8.47$ & $9.14$ & $8.65$ & $34424$ & $33570$ & $33896$\\
				Model 3$^{\text{1}}$ & $6.73$ & $9.22$ & $7.06$ & $35146$ & $44772${*} & $33668$\\
				Model 3$^{\text{2}}$ & $5.61$ & $7.83$ & $5.89$ & $31566$ & $46602${**} & $30370$\\
				Model 4$^{\text{1}}$ & $7.45$ & $8.02$ & $7.61$ & $34016$ & $31954$ & $33338$\\
				Model 4$^{\text{2}}$ & $5.39$ & $5.77$ & $5.49$ & $26287$ & $27603$ & $26114$\\
				Model 5$^{\text{1}}$ & $4.60$ & $4.64$ & $4.61$ & $33712$ & $34230$ & $33675$\\
				Model 5$^{\text{2}}$ & $8.99$ & $12.06$ & $9.43$ & $30544$ & $21631$ & $29239$\\
				\hline
			\end{tabular}
			\tablefoot{$^{\text{1}}$: primary star, $^{\text{2}}$: secondary star.	*\&**: These values are not physically acceptable due to the problems in treating the reflection effect near L$_1$, the most probable values based on the temperature of neighbouring grid points are: *: $27000$\,K; **: $26300$\,K.}
	\end{table*}

	\begin{table*}
		\caption{Gravity (in cgs units).}
		\label{tab:Gravity}
		\centering
			\begin{tabular}{l c c c c c c}
				\hline\hline
				\small
				Stars & $\log(g(R_{\mathrm{min}}))$ & $\log(g(R_{\mathrm{max}}))$ & $\overline{\log(g)}$ &
				$\log\left(\dfrac{GM}{R_{\mathrm{min}}^{2}}\right)$ & $\log\left(\dfrac{GM}{R_{\mathrm{max}}^{2}}\right)$ &			$\log\left(\dfrac{GM}{R_{\mathrm{mean}}^{2}}\right)$\\[2.5mm]
				\hline
				\normalsize 
				Model 1$^{\text{1}}$ & $4.03$ & $3.95$ & $4.01$ & $4.02$ & $3.99$ & $4.01$\\
				Model 1$^{\text{2}}$ & $4.05$ & $3.98$ & $4.03$ & $4.05$ & $4.01$ & $4.04$\\
				Model 2$^{\text{1}}$ & $3.93$ & $3.78$ & $3.89$ & $3.92$ & $3.85$ & $3.90$\\
				Model 2$^{\text{2}}$ & $3.93$ & $3.78$ & $3.89$ & $3.92$ & $3.85$ & $3.90$\\
				Model 3$^{\text{1}}$ & $4.13$ & $2.67$ & $4.05$ & $4.12$ & $3.84$ & $4.08$\\
				Model 3$^{\text{2}}$ & $4.13$ & $2.67$ & $4.04$ & $4.11$ & $3.82$ & $4.06$\\
				Model 4$^{\text{1}}$ & $3.96$ & $3.81$ & $3.92$ & $3.95$ & $3.88$ & $3.93$\\
				Model 4$^{\text{2}}$ & $3.98$ & $3.85$ & $3.95$ & $3.97$ & $3.91$ & $3.96$\\
				Model 5$^{\text{1}}$ & $4.31$ & $4.29$ & $4.30$ & $4.31$ & $4.30$ & $4.30$\\
				Model 5$^{\text{2}}$ & $3.53$ & $2.62$ & $3.45$ & $3.51$ & $3.26$ & $3.47$\\
				\hline
			\end{tabular}
			\tablefoot{Columns two to four indicate the extreme values and the average of the actual $\log(g)$ on the deformed stellar surface accounting for the Roche potential, whilst the last three columns provide the corresponding $\log(g)$ computed with the single star formula.}
	\end{table*}

	The most compact systems, Model 1, Model 2, Model 4 and the primary star of Model 5, present small variations of their radius, $\log(g)$ and temperature. The agreement between the single star gravity and the gradient of the Roche potential is quite good. However, for Model 3 and the secondary of Model 5, the differences are more important. For Model 3, the difference between the maximum and minimum radius is about $2.5R_{\sun}$ for the primary and $2.2R_{\sun}$ for the secondary, $\log(g)$ varies from $4.1$ to $2.7$ for both stars (Fig. \ref{fig:log(g)Model3} shows $\log(g)$ variations for the Model 3). These variations of the surface gravity lead to a change of temperature by about $8000$ K for the primary star and $5000$ K for the secondary star (Fig. \ref{fig:tempvarModel3} shows temperature variations for the Model 3). The difference between the single star $\log(g)$ and the binary model $\log(g)$ is $1.4$. However, this difference appears only for the most deformed part of the star. At the poles, the difference between the single star model and binary model is negligible. For the secondary star of Model 5, the difference between the maximum and minimum radius is $3R_{\sun}$, $\log(g)$ varies from $3.5$ to $2.6$. These variations of the surface gravity lead to a change of temperature by about $8000$ K. The difference between the single star $\log(g)$ and the binary system value is $1.24$. Again, this difference concerns only the most deformed part of the star. Near the poles, the difference between the two models is negligible. 
	
	We also test a disentangling technique on our combined spectra to compare the output of this method with the true spectra of the binary components in our simulations. For this purpose, we use the Gonz\'alez \& Levato (\cite{Gonzalez}) disentangling technique as implemented by Linder et al. (\cite{Linder}) and compare the result to the individual spectra computed with our method. The results show a very good agreement between disentangled spectra and computed ones. 
	
	We now consider two problems encountered in observational studies of massive binaries: the spectral classification and the Struve--Sahade effect.

	\subsection{Spectral classification}

		First, we use the quantitative Conti (\cite{Conti})--Mathys (\cite{Mathys(1)}, \cite{Mathys(2)}) criterion. This criterion is based on the ratio between the equivalent widths (EWs) of He\,I $\lambda$\,4471 and He\,II $\lambda$\,4542 for the spectral type determination and on the ratio between the EWs of Si\,IV $\lambda$\,4089 and He\,I $\lambda$\,4143 for luminosity class determination. The spectral types that we obtain for our simulated spectra with this criterion are in good agreement with those obtained by the observational studies of the corresponding binary systems. However, the luminosity classes are very different. The criterion applied to simulated spectra indicates supergiants whilst observations indicate main-sequence stars (Table \ref{tab:SpecClassCM} gives the classification of the stars with the Conti--Mathys criterion). The problem seems to come from the He\,I $\lambda$\,4143 line. This line is a singlet transition of He\,I. Najarro et al. (\cite{Najarro}) reported pumping effects between He\,I and Fe\,IV lines that are not properly taken into account in model atmosphere codes and lead to wrong He\,I line strengths for singlet transitions. To classify our synthetic stars, this line is therefore not well suited and we use the Walborn \& Fitzpatrick (\cite{Walborn}) atlas instead. This atlas is more qualitative than the Conti--Mathys criterion but allows us to take account of a larger wavelength domain to classify the stars. The Conti--Mathys criterion has been evaluated on the combined spectrum of the system, on the individual spectra of  each component and on the disentangled spectra. The disentangling is performed on 20 simulated spectra sampling the full orbital cycle. The classification based on the Walborn \& Fitzpatrick (\cite{Walborn}) atlas has been done on individual synthetic spectra. For observational data, we have to disentangle the spectra to use the atlas.		
		The atlas of Walborn \& Fitzpatrick (\cite{Walborn}) draws attention on many lines such as He\,I $\lambda$\,4471, He\,I $\lambda$\,4387, He\,II $\lambda$\,4542, Si\,IV $\lambda$\,4088 and He\,II $\lambda$\,4200. By observing these lines (and their relative depth) and comparing our spectra to the references of the atlas we reclassify the stars (see Table \ref{tab:SpecClassWF}) achieving now a better agreement with the classification of the real binary systems.

		\begin{table*}[ht!]
			\caption{Spectral classification using the Conti--Mathys criterion.}
			\label{tab:SpecClassCM}
			\centering
				\begin{tabular}{l c c c c}
					\hline\hline
					Stars & Combined spectra & Separated spectra & Disentangled spectra & 
					Observational analysis\\
					\hline 
					Model 1$^{\text{1}}$ & O6.5I & O6.5I & O6.5I & O7V\\
					Model 1$^{\text{2}}$ & O6.5I & O6.5I & O6.5I & O7.5V\\
					Model 2$^{\text{1}}$ & O7.5I & O7.5I & O7.5I & O6V\\
					Model 2$^{\text{2}}$ & O7.5I & O8I & O8III & O6.5V\\
					Model 3$^{\text{1}}$ & O8.5V & O8.5V & O8.5V & O7.5V\\
					Model 3$^{\text{2}}$ & O9.5V & O9.7V & O9.7V & O9.5V\\
					Model 4$^{\text{1}}$ & O8.5I & O8.5III & O8.5III & O9III (or V)\\
					Model 4$^{\text{2}}$ & B-star & B-star & B-star & B1III (or V)\\
					Model 5$^{\text{1}}$ & O9III & O9V & O9V & O8III\\
					Model 5$^{\text{2}}$ & O9.5V & O9.5V & O9.5V & O9.5III\\
					\hline
				\end{tabular}
		\end{table*}

		\begin{table}
			\caption{Spectral classification using the Walborn \& Fitzpatrick (1996) atlas }
			\label{tab:SpecClassWF}
			\centering
				\begin{tabular}{l c c}
					\hline\hline 
					Stars & Synthetic spectra & Observed spectra\\
					\hline 
					Model 1$^{\text{1}}$ & O6.5-O7V & O7V\\
					Model 1$^{\text{2}}$ & O7-O7.5V & O7V\\
					Model 2$^{\text{1}}$ & O7-O7.5V (or III) & O7V\\
					Model 2$^{\text{2}}$ & O8-O8.5V & O7.5-O8V\\
					Model 3$^{\text{1}}$ & O8-O8.5V (or III) & O7.5V\\
					Model 3$^{\text{2}}$ & O9.5-O9.7V (or III) & O9.5V\\
					Model 4$^{\text{1}}$ & O8.5-O8V (or III) & O9V\\
					Model 4$^{\text{2}}$ & B0.5-B0.2V & B1V\\
					Model 5$^{\text{1}}$ & O8.5-O9III (or V) & O9V\\
					Model 5$^{\text{2}}$ & O9.5-O9.7III (or V) & O9.5III\\
					\hline
				\end{tabular}
		\end{table}
		
	\subsection{Struve--Sahade effect }

		The Struve--Sahade effect is usually defined as the apparent strengthening of the secondary spectrum when the star is approaching the observer and weakening when it moves away (Linder et al., \cite{Linder}). This effect was first reported by Struve (\cite{Struve}). He proposed that this was due to streams of gas moving to the secondary and obscuring it. Gies, Bagnuolo \& Penny (\cite{Gies}) rather suggested that the phenomenon might be due to heating by back scattering of photons by the stellar wind interaction zone. Stickland (\cite{Stickland}) proposed that the Struve--Sahade effect could be linked to a combination of several mechanisms. Gayley (\cite{Gayley}) and Gayley et al. (\cite{Gayleyetal}) argued that this effect could be linked to flows at the surface of the stars. For a more extensive discussion we refer the reader to Linder et al. (\cite{Linder}). In our sample of binaries, three are known to present this effect: HD\,159176, HD\,165052, HD\,100213 that respectively inspired our Model 1, Model 2, Model 3.

		As a first step, we investigate the variation of some lines in the spectra at 20 different phases to look for variations of the profile. Especially, we inspect the extreme phases $0.25$ and $0.75$ where the two components are the most separated to see differences in the relative strength of the lines. The result of this qualitative examination is given in Table\,\ref{tab:VisSS}. We must pay attention to possible blends that can modify the strength of a line leading to an apparent Struve--Sahade effect (see below).

		\begin{table*}[ht!]
			\caption{Visual variations of line intensity.}
			\label{tab:VisSS}
			\centering
				\begin{tabular}{l c c c c c}
					\hline\hline
					Line & Model 1 & Model 2 & Model 3 & Model 4 & Model 5\\
					\hline 
					He\,I $\lambda$\,4026 & 1 & 1b & 2 & 2 & 2\\
					He\,II $\lambda$\,4200 & 2 & 2 & / & 1B & 2\\
					He\,I $\lambda$\,4471 & 1bs & 1B & 1b & 2 & 1B\\
					He\,II $\lambda$\,4542 & 1s & 1sB & 1B & 2 & 1B\\
					He\,II $\lambda$\,4686 & 2 & 2 & 2 & 2 & 1B\\
					He\,I $\lambda$\,4713 & 1B & 1sB & 1B & 1B & 1B\\
					H\,$\beta$ & 1b & 1B & 1sb & 2 & 2\\
					He\,I $\lambda$\,4921 & 1B & 1B & 2 & 2 & 2\\
					He\,I $\lambda$\,5016 & 1 & 2 & 1B & 2 & 1b\\
					He\,II $\lambda$\,5412 & 1s & 2 & 2 & 2 & 2\\
					C\,III $\lambda$\,5696 & 1B & 1B & 2 & 2 & 2\\
					C\,IV $\lambda\lambda$\,5801, 5812 & 2 & 2 & 2 & 2 & 2\\
					He\,I $\lambda$\,5876 & 2 & 2 & 2 & 2 & 2\\
					\hline
				\end{tabular}
				\tablefoot{1: variation, 2: no variation, b: suspicion of blend or small blend in the wings, B: blend with other lines, s: small variations.}
		\end{table*}
		
		\begin{figure}[ht!]
			\centering
			\includegraphics[height=7.5cm, width=6cm]{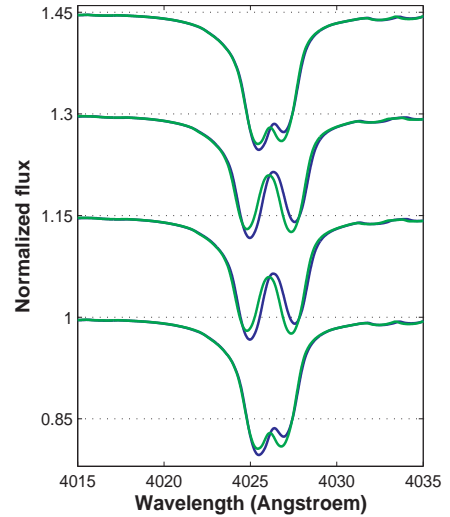}
			\caption{Variations of the He\,I $\lambda$\,4026 line in Model 2. From bottom to top, in blue: phases from $0.1$ to $0.4$ by step of $0.1$. From top to bottom, in green: phases from $0.6$ to $0.9$ by step of $0.1$.}
			\label{fig:HeI4026varModel2}
		\end{figure}
		
		First of all, we detect quite a lot of variations in the lines. However, many of these variations can be explained by blends with weak nearby lines rather than true variations of the line intensities. Let us stress that, a priori, we do not expect our simulations to display a genuine Struve--Sahade effect, because the simulations are axisymmetric about the binary axis and the largest difference due to non-uniform surface gravity and temperature are thus expected when comparing phases near $0.0$ and $0.5$, rather than $0.25$ and $0.75$ which are the phases most concerned by the Struve--Sahade effect (see for example Fig. \ref{fig:HeI4026varModel2}). We have also performed a quantitative study on several lines to evaluate the real variations. In order to check the influence of a possible blend, we measure the equivalent width on the simulated combined spectra of the binary at different phases as well as on the spectra of the individual components of the binary at the same phases. The measurements of the line EWs are done with the MIDAS software developed by ESO. The EWs of lines in the spectra of the individual stars are determined directly by simple integration with the integrate/line routine whilst for the binary system, we use the deblend/line command as we would do for actual observations of binary system. The latter routine fits two Gaussian line profiles to the blend of the primary and secondary lines. We stress again that for real observations, we do not have access to the individual spectra of the primary and secondary at the various orbital phases. Our simulations allow us to compare the ``observed" EWs (those measured by deblend/line) to the ``real" ones (those measured on the individual spectra with integrate/line) and to distinguish real variations from apparent variations due to the biases of the deblending procedure. We now consider the simulations for the various systems listed in Table \ref{tab:*PARAM}.

		\subsubsection{Model 1}

\begin{figure}
				\resizebox{\hsize}{!}{\includegraphics{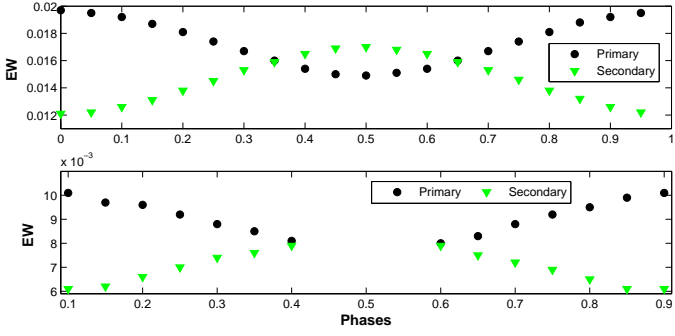}}
				\caption{EWs (in \AA) of C\,III $\lambda$\,5696 of Model 1. 
				Upper panel: EWs of primary and secondary measured on individual spectra. 
				Lower panel: EWs of primary and secondary measured on combined spectra.}
				\label{fig:EW(CIII)}
			\end{figure}

			We first study individual spectra of each component. The lines listed in Table \ref{tab:VisSS} present small variations of $1-3$\% over the orbital cycle. These variations are phase-locked. However, two lines present larger variation of more than $20$\%: He\,I $\lambda$\,4921 and C\,III $\lambda$\,5696 (predicted in photospheric emission \footnote{The apparent inverted P-Cygni profile is due to the proximity of an absorbtion Al\,III unrelated to the C\,III line.}). 
			Unfortunately, the He\,I $\lambda$\,4921 line is a singlet transition of He\,I and is thus subject to the effect described by Najarro et al. (2006) who showed that the singlet helium transitions are, in atmosphere codes like TLUSTY, very sensitive to the treatment of the spectrum of Fe\,IV.	
			The variation of the second line is more difficult to explain.
			If there effectively exist a small blend on the blue component, the blended line is very weak and it seems improbable that this small blend can explain $25-30$\% of variations.
			The variations on the combined spectra are more difficult to see. Furthermore, the C\,III line is the only one that presents large variations on the combined spectra (see Fig. \ref{fig:EW(CIII)}). We stress here that this line is rather easy to measure in our simulated data: the lines of the primary and secondary are very well separated, the continuum is well defined. The results are then more reliable than for other lines. The variations are phase-locked and the lines are stronger when the stars present their hotter face. However, the predicted strength of the C\,III lines are rather weak and would make a measurement on real data (with noise) quite difficult.
			
			The primary/secondary EW ratios (all EW ratios in this paper have been measured on combined spectra with the deblend/line routine) also present variations such as these shown on Fig. \ref{fig:EW_ratio_He4026} except for three lines. This pattern of variations appears in other models and we call it Pattern 1.
			He\,I $\lambda$\,4713 presents an inverse V variation pattern with a peak centered on phase $0.5$. He\,I $\lambda$\,4921 presents irregular variations with large differences between the first and the second half of the orbital cycle. And finally, C\,III $\lambda$\,5696 presents a V-shape variation which is expected in regard of previous results (see Fig. \ref{fig:EW(CIII)}).

		\subsubsection{Model 2}
		
\begin{figure}
				\resizebox{\hsize}{!}{\includegraphics{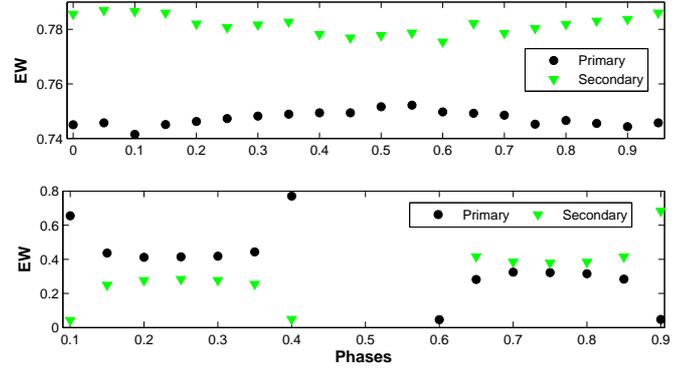}}
				\caption{ Same as Fig.\ref{fig:EW(CIII)} for the He\,I $\lambda$\,4026 line in Model 2.}
				\label{fig:EW_HeI4026}
			\end{figure}

			First we study the individual spectra of each component. We can see variations of intensities for all the lines. However, in most cases, these variations are very small of the order of $1-2$\% and will probably escape detection in real observations. The variations seem phase-locked as we can see on Fig.\ref{fig:EW_HeI4026} and reflect the temperature over the visible part of the stellar surface.

			The analyses on the combined spectrum suffer from the same difficulties as for Model 1, and are sometimes in opposition with what we observe on the individual simulated spectra: the EW of the primary line is always lower than that of the secondary while the combined spectrum shows the primary line to be deeper than the secondary at phases $0.0$ to $0.5$ (e.g. He\,I $\lambda$\,4026, He\,I $\lambda$\,4471, see Fig. \ref{fig:EW_HeI4026}).

			The He\,I $\lambda$\,4921 line displays a more important variation of the order of $15$\% (see however the remark about this line in the discussion of Model 1). Moreover, on the combined spectrum, this line is blended with an O\,II line at $\lambda$\,4924.5 so it is very difficult to determine whether the variation comes from the blend or from another effect.
			
\begin{figure}[ht!]
				\centering
				\includegraphics[width=6cm]{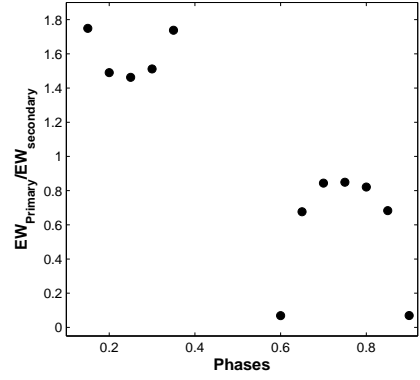}
				\caption{EW ratio between primary and secondary for He\,I $\lambda$\,4026 in Model 2.}
				\label{fig:EW_ratio_He4026} 
			\end{figure}

			A last point is the study of the ratio between the EWs of primary and secondary lines. The results present Pattern 1 variations (see Fig.\ref{fig:EW_ratio_He4026}).

\begin{figure}
				\resizebox{\hsize}{!}{\includegraphics{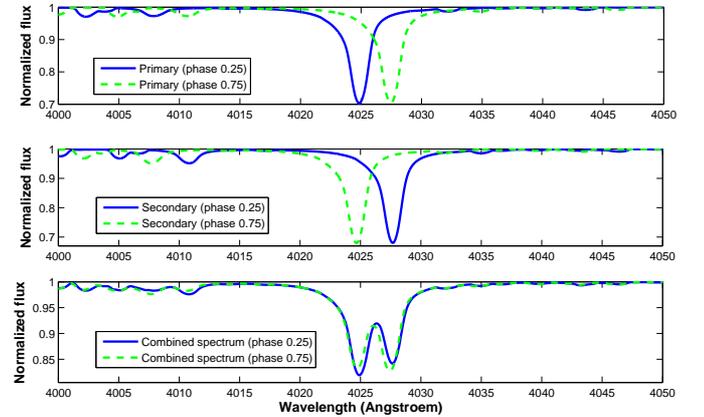}}
				\caption{He\,I $\lambda$\,4026 line at phases $0.25$ and $0.75$ in Model 2.
					 Upper panel: Primary spectra.
					 Middle panel: secondary spectra. 
					 Lower panel: combined spectra.}
				\label{fig:HeI4026}		
			\end{figure}

			Fig.\ref{fig:HeI4026} illustrates the He\,I $\lambda$\,4026 line at two opposite phases. Our individual spectra clearly reveal this line to be asymmetric with a steeper red wing for both stars. When fitting the combined binary spectra with two symmetric Gaussian profiles, the red component will be systematically assigned a lower flux, because the blue component will be apparently broader than the red one. Although the resulting fit can be of excellent quality, it will provide systematically larger EWs for the blue and lower EWs for the red component. This situation therefore leads to an artificial Struve--Sahade behaviour.
			
		\subsubsection{Model 3}
		
\begin{figure}
				\resizebox{\hsize}{!}{\includegraphics{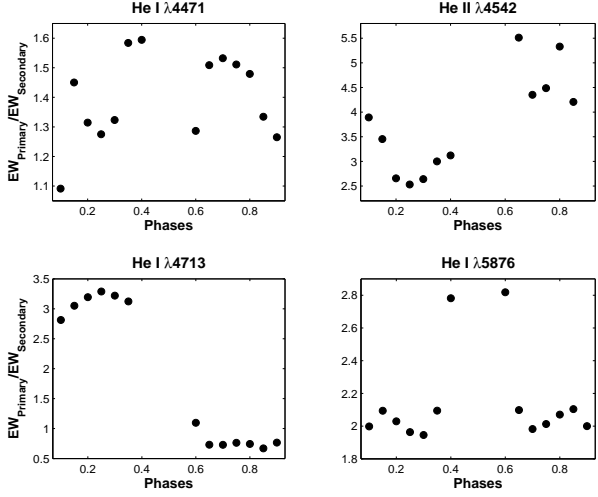}}
				\caption{EW ratio between primary over secondary for several lines (from left to right, top to bottom: He\,I $\lambda$\,4471, He\,II $\lambda$\,4542, He\,I $\lambda$\,4713, He\,I $\lambda$\,5876) in Model 3.}
				\label{fig:EW_ratio_Model3}
			\end{figure}

			The analysis of Model 3 is more complex than for the two other systems. The results show different behaviours.
			 The EW ratios of the primary and secondary measured on combined spectra present different variation patterns (see Fig. \ref{fig:EW_ratio_Model3}). The He\,I $\lambda$\,4471 line presents Pattern 1-like variations. For the first part of the orbital cycle, He\,II $\lambda$\,4542, He\,I $\lambda$\,4713 and He\,I $\lambda$\,5016 lines show variations that seem phase-locked. However, the second part of the orbital cycle is either constant or irregular. Finally, the He\,I $\lambda$\,5876 line shows phase-locked variations, symmetric with respect to phase $0.5$. The variations have an W-shape pattern (see Fig. \ref{fig:EW_ratio_Model3}).	 
			 
			The EWs of the He\,I $\lambda$\,4471 line for the separated spectra of the primary show a maximum at phase $0.5$ when the star presents its coolest face (the front part is never visible because of the eclipse). Conversely, the secondary shows a minimum at this phase when only the hottest parts of the star are visible (the coolest are occulted by the primary). The variations are of the order of $10$\%. The analysis of the combined spectra is more complex. For the first half of the orbital cycle, the EWs of the primary star have a similar behaviour to the measurements on individual spectra. For the second part of the orbital cycle and for the secondary star, the variations do not seem phase-locked.
			The EWs of the He\,II $\lambda$\,4542 line for the separated spectra show a V-shape pattern for the primary and an inverse V-shape pattern for the secondary. The variations are of the order of $20$\% and phase-locked. The maximum (resp. minimum) of the curves occurs when the stars present their hottest (resp. coolest) face. Again, the study of the combined spectra shows irregular variations of the same order of magnitude as the study of the separated spectra. 
			For the He\,I $\lambda$\,4713 and He\,I $\lambda$\,5016 lines, the proximity of other lines renders the analysis more complex because of blends. The variations seem irregular except for the He\,I $\lambda$\,5016 line in the separated spectra of the primary star which shows an inverse V-shape pattern.
			Finally, the EWs of the He\,I $\lambda$\,5876 line for the separated spectra of the primary are constant. The EWs for the separated spectra of the secondary are more interesting and show variations of about $10$\% and an inverse V-shape pattern with a peak at phases near $0.5$. The EWs of the combined spectra of the primary present a maximum peaked at phase $0.5$ and are quite constant over the rest of the orbital cycle. The secondary shows a minimum at phase $0.5$ and small variations at other phases.
			As we see, this system is more complex than the other two. The variations are more important but often symmetric with respect to phase $0.5$. They seem also phase-locked and more important than for the other stars (for the individual spectra at least). 
			
	\subsection{Broad-band light curves }
		
\begin{figure}
			\resizebox{\hsize}{!}{\includegraphics{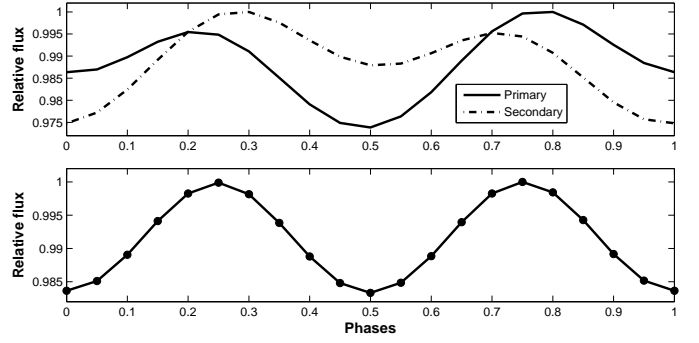}}
			\caption{Synthetic light curves of Model 1 in the range $\lambda\lambda$ 3500, 7100.
			Upper panel: Light curves of the primary and secondary stars (normalized to the maximum of emission).
			Lower panel: Combined light curve of the binary system (normalized to the maximum of emission).}
			\label{fig:LightModel1}
		\end{figure}

\begin{figure}
			\resizebox{\hsize}{!}{\includegraphics{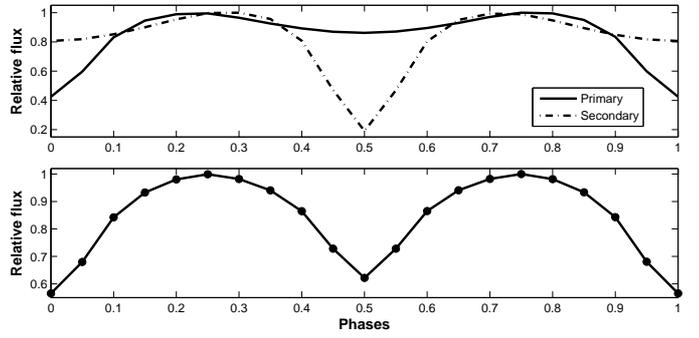}}
			\caption{Same as Fig.\ref{fig:LightModel1}, but for Model 3.}
			\label{fig:LightModel3}
		\end{figure}

		The analysis of the broad-band light curves is quite interesting because we can see variations of the intensity over the orbital cycle. Of course, in case of an eclipsing binary, the light curve is dominated by the eclipses (Fig.\ref{fig:LightModel3}). However, even for non eclipsing systems such as Model 1, variations occur as we can see in Fig.\ref{fig:LightModel1}. These feature are the so-called ellipsoidal variations that are well known. Moreover, an interesting result is the asymmetry of the curve. In the first half of the orbital cycle, the primary star is fainter than in the second half. Conversely, the secondary star is brighter in the first half and fainter in the second half. Adding both light curves up results in a roughly symmetrical light curve for this particular system.
		
		Indeed, over the first half of the orbital cycle, the primary star has a negative radial velocity and it spectral energy distribution is Doppler shifted to the blue whilst that of the secondary is red-shifted. The behaviour of the light curves can indeed be explained by the displacement of the emission peak of the continuum. The difference between the flux emitted over the wavelength range $\lambda\lambda$ 4000,6000 \AA\ by an O9 star at rest and by an O9 star with a radial velocity of $100$ kms$^{-1}$ is about $1$\%. Therefore, the shift of the maximum peak of emission towards the blue (resp. red) in the first (resp. second) half of the orbital cycle for the primary (resp. secondary) and conversely for the second half of the orbital cycle induces the asymmetries in the broad-band light curve.

	\subsection{Radial velocity curve }

		The last aspect that we analyze is the radial velocity (RV) curve. As we know the velocity of each visible point at a given phase, we can compute the mean value of these points and thus obtain a mean radial velocity. The curves are symmetric for all the five systems and the semi-amplitudes (K) are in good agreement with the observed ones. This shows that the lines that were used in the analyses of the observations represent the true RV of the star rather well. Table \ref{tab:RV} gives a comparison between observed and simulated semi-amplitudes of the radial velocity curves.

		\begin{table}
			\caption{Comparison between observed and computed radial velocity semi-amplitudes.}
			\label{tab:RV}
			\centering
				\begin{tabular}{l c c}
					\hline\hline
					Stars & Computed K & Observed K \\
					 & (kms$^{-1}$) & (kms$^{-1}$)\\
					\hline 
					Model 1$^{\text{1}}$ & $209.1$ & $210.3$\\
					Model 1$^{\text{2}}$ & $218.0$ & $216.5$\\
					Model 2$^{\text{1}}$ & $97.1$ & $96.4$  \\
					Model 2$^{\text{2}}$ & $112.1$ & $113.5$ \\
					Model 3$^{\text{1}}$ & $249.5$ & $249.7$   \\
					Model 3$^{\text{2}}$ & $368.4$ & $369.2$   \\
					Model 4$^{\text{1}}$ & $143.0$ & $167.1$\\
					Model 4$^{\text{2}}$ & $300.8$ & $301.3$\\
					Model 5$^{\text{1}}$ & $143.0$ & $143.7$\\
					Model 5$^{\text{2}}$ & $230.9$ & $230.6$\\
					\hline
				\end{tabular}
		\end{table}

		A non-uniform surface temperature distribution in HD\,100213 has been reported by Linder et al. (\cite{Linder}), so we investigate more in detail our Model 3 to check whether there exists such a distribution in our simulation. We study the temperature distribution by computing the amplitude of the radial velocity curves of He\,I and He\,II lines. As we can see in Table \ref{tab:RVModel3} the agreement between the model and the observation is not very good. The observations suggest that the He\,I lines are formed over the rear part of the stars (because of larger K values). This region then has to correspond to the lowest temperatures. However, in our model, the lowest temperatures are localized both in front and in rear parts of the stars. Hence our model shows an intermediate amplitude that suggests a mean between front and rear formation regions.
		
		For the He\,II lines, the observations suggest that they are formed in the intermediate and lower radial velocity amplitude region. Thus, the lines are formed over the front and middle parts of the star. However, our model shows that the lines are only formed in the middle part of the star and not in the front.
		Thus, we can conclude that there probably exists an additional mechanism that enhances the reflection process and heats the front part of the stars. This mechanism could be linked to the stellar wind interaction that might warm the stellar surface. 
		
		In Sect. 2.3, we have mentioned the problem of the increase of the temperature near the L1 point in the reflection treatment. We stress that this increase can, a priori, not be the mechanism that heats the front parts of the stars. Indeed, it is really a problem of convergence affecting very few points. For Model 3, after $50$ iterations, the model was not yet converged and the temperature of the problematic points ($3$ points over $14162$ for this model) were at more than $100$ kK. In comparison, the convergence for the non-problematic points is reached within less than 5 iterations.

		\begin{table}[ht!]
			\caption{Semi-amplitudes of the radial velocity curves (in kms$^{-1}$) of the He lines in the Model 3 spectra.}
			\label{tab:RVModel3}
			\centering
				\begin{tabular}{l c c c c c c}
					\hline\hline  
					Lines & K$_1$ &  K$_2$ & K$_{1\mathrm{, m}}$ &  K$_{2\mathrm{, m}}$ & K$_{1\mathrm{, obs}}$ &  K$_{2\mathrm{, obs}}$\\
					\hline
					He\,I $\lambda\,4026$ & $245.0$ & $354.4$ & $223.4$ & $372.3$ & $246.5$ & $363.8$ \\
					He\,I $\lambda\,4471$ & $236.0$ & $353.7$ & $234.7$ & $368.8$ & $244.5$ & $372.3$ \\
					He\,II $\lambda\,4686$ & $265.5$ & $364.4$ & $255.9$ & $383.9$ & $251.3$ & $355.2$ \\
					He\,I $\lambda\,4921$ & $284.5$ & $364.9$ & $255.1$ & $365.5$ & $245.9$ & $373.0$ \\
					He\,II $\lambda\,5412$ & $233.5$ & $332.4$ & $249.3$ & $360.1$  & $239.8$ & $351.2$ \\
					He\,I $\lambda\,5876$ & $247.9$ & $354.6$ & $229.6$ & $357.2$  & $269.7$ & $388.8$ \\
					\hline 
				\end{tabular}
				\tablefoot{The first two columns are computed from the measurement of the Doppler shift of the center of the lines on the separated spectra of the primary and the secondary, the next two are the results obtained with the MIDAS (deblend/line) measurement of the center of the lines and the last two are the semi-amplitudes observed by Linder et al.(\cite{Linder}) for HD\,100213.}
		\end{table}

\section{Summary and future perspectives}

	In this paper we have presented a mathematical model, based on first principles, that allows to compute the physical properties on the surface of stars in circular orbit massive binary systems featuring main-sequence O stars. We assume that the stars are corotating with their orbital motion, allowing us to use the Roche potential formalism to infer the shape of the stars and the local acceleration of gravity. We include gravity darkening and reflection effects to infer the surface temperature distribution of the stars. These results are then combined with the TLUSTY model atmosphere code to compute synthetic spectra of each binary component and of the combined binary as a function of  orbital phase. Many of the spectral lines are found to display phase-locked profile and/or intensity variations in our simulated spectra. These variations are quite small for well detached systems, but their amplitude strongly increases for heavily deformed stars. This variability mostly reflects the non-uniform temperature distribution over the stellar surface which is seen under different orientations as a function of orbital phase. Given the assumptions that we have made in our model, we expect these variations a priori to be symmetric with respect to the conjunction phases. Our simulations demonstrate the impact of line blending on the measurement of lines in massive binary systems. Indeed, for a large number of lines, we find that the true variations of the EWs (symmetric with respect to the conjunction phases) differ from those that are measured on the combined binary spectra using conventional deblending methods and which are often found to be asymmetric with respect to the conjunction phases. The most dramatic effect is seen when the intrinsic line profiles of the individual stars are asymmetric (e.g. due to blends with weaker lines in the wings of a strong line). In this case, deblending the binary spectra with some type of symmetric line profile (Gaussian or Lorentzian) introduces a systematic effect that mimics the so-called Struve--Sahade effect. This result hence brings up an alternative explanation for the Struve--Sahade effect: in some binary systems and at least for some lines, this effect could simply be an artifact due to the fit of blended asymmetric lines with symmetric profiles. Therefore, at least in some cases, the Struve--Sahade effect does not stem from genuine physical processes, but rather reflects a bias in the measurement of the line profiles.
	
	Whilst deblending appears problematic, our simulations demonstrate that spectral disentangling, assuming a reasonable sampling of the orbital cycle, yields good results, that provide a good representation of the average spectrum of a binary component and can hence be used for spectral classification and model atmosphere fitting. Comparing the RV amplitudes measured for different lines in the spectrum of the contact binary system HD100213 with the predicted RV amplitudes for this system, we found that our model apparently fails to reproduce the surface temperature distribution of this system. Additional heating of those parts of the stars that face each other would be needed to account for the observed distribution.
	
As the next steps, we intend to generalize our code in different ways. For instance, we plan to extend our approach to more evolved stars that feature extended atmospheres. This requires both a revision of the Roche potential formalism to account for the effect of radiation pressure (see e.g. Howarth, \cite{Howarth}) and the usage of a model atmosphere code that includes the effect of a stellar wind. Another important aspect is to adapt our code to asynchronous and/or eccentric binary systems. In these systems, the Roche potential formalism is no longer valid and must be replaced by a proper handling of tidal effects (see e.g. Moreno, Koenigsberger, \& Harrington \cite{Moreno}). Finally, we plan to account for the presence of a wind interaction zone between the stars. The latter can contribute to the heating of the stellar surface in two different ways, either by backscattering of the photospheric photons or by irradiation with X-ray  photons emitted by the shock-heated plasma in the wind interaction zone. These improvements will be the  subject of forthcoming papers.

\begin{acknowledgements}
We acknowledge support through the XMM/INTEGRAL PRODEX contract (Belspo), from the Fonds de Recherche Scientifique (FRS/FNRS), as well as by the Communaut\'e Fran\c caise de Belgique - Action de recherche concert\'ee - Acad\'emie Wallonie - Europe.
\end{acknowledgements}


\end{document}